\documentstyle[manuscript,prl,aps]{revtex}
\begin{document}
\title{Connection between Calogero-Marchioro-Wolfes type few-body models and
free oscillators}
\author{N. Gurappa$^1$, Avinash Khare$^{2,\dagger}$, and Prasanta. K.
Panigrahi$^{1}$
\thanks{panisp@uohyd.ernet.in , $\dagger$ khare@iopb.stpbh.soft.net}} 
\address{1. School of Physics, University of Hyderabad, Hyderabad 500 046,
India\\ and\\
2. Institute of Physics, Sachivalaya Marg, Bhubaneswar 751 005, India}

\maketitle

\begin{abstract} 
We establish the exact correspondence of the Calogero-Marchioro-Wolfes model
and several of its generalizations with free oscillators. This connection
yields the eigenstates and leads to a proof of the quantum integrability. The
usefulness of our method for finding new solvable models is then
demonstrated by an example.
\end{abstract}
\draft

\newpage
Exactly solvable, many-body, interacting systems in one and higher dimensions
constitute one of the most exciting branch of active research in physics.
Three decades ago, Calogero formulated and solved the quantum mechanics of
three identical particles in one-dimension, interacting via pair-wise
harmonic and inverse-square potentials [1] and subsequently generalized it to
$N$ particles [2]. The connection of this model with random matrices was
established by Sutherland who also analysed its thermodynamic behavior [3].
This and related models, popularly known as Calogero-Sutherland (CS) models,
have found interesting applications in diverse branches of physics [4].
Models with more than two-body interactions have also attracted considerable
attention in the literature. Wolfes has solved the three particle Calogero
system in the presence of a three-body interaction of the type $\sum_{i\ne
{j\ne {k\ne i}}}^3 (x_i + x_j - 2 x_k)^{-2}$ [5]; Calogero and Marchioro
pointed out the novel aspects of the scattering problem [6]. A number of
generalizations of this Calogero-Marchioro-Wolfes (CMW) model have been
recently obtained using supersymmetric quantum mechanics [7]. Further, using
exchange operator formalism, the algebra of the CMW model has been shown to
be $D_6$-{\it extended} Heisenberg algebra [8]. Note that, for pure two-body
and three-body cases, the corresponding symmetries are $S_3$ and $D_3$-{\it
extended} Heisenberg algebras respectively. The striking features of many of
these models are the harmonic oscillator type excited spectra and the
coupling dependence of the ground-state energy.

Recently, two of us have shown that the $N$-particle CS models can be mapped
identically to free harmonic oscillators by a similarity transformation (ST)
[9,10]. The fact that the energy spectra of both CS and CMW models are
identical in their structure motivates us to search for a possible mapping of
the CMW type models to free harmonic oscillators. The purpose of this note is
to show that such an exact mapping indeed exists. The importance of this
result lies in the fact that, it naturally explains why CMW type models have
a linear spectra and other properties like quantum integrability. This
equivalence leads to the explicit construction of the wavefunctions, starting
from the symmetrized eigenfunctions of the free harmonic oscillators.
Furthermore, it paves the way for non-trivial generalizations of the CMW
model to more than three particles. We demonstrate this fact by explicitly
constructing a new, solvable four particle interacting model which is quantum
integrable. Further, we consider several 3-body problems with oscillator-like
spectrum and show that all of them can be mapped to a set of decoupled
oscillators.

We begin with the Calogero-Marchioro-Wolfes system given by the Hamiltonian
($\hbar = \omega/2 = 2 m = 1$)  
\begin{equation}
H = - \sum_{i=1}^3 {\partial}_i^2
+ \sum_{i=1}^3 x_i^2 + g \sum_{{i,j=1}\atop {i\ne j}}^3 \frac {1}{x_{ij}^2} +
3 f \sum_{{i,j=1}\atop {i\ne j}}^3 \frac {1}{y_{ij}^2}\qquad, 
\end{equation}
here, ${\partial}_i \equiv \frac{\partial}{\partial x_i}$, $x_{ij}
\equiv x_i - x_j, y_{ij} \equiv x_i + x_j - 2 x_k$; $i \ne j \ne k \ne i$ and
$g > - \frac{1}{4} < f$ are the coupling constants. 

The ground-state wavefunction is given by
\begin{equation}
\psi_0 = G\,\,|X|^\alpha\,\,|Y|^\lambda \qquad,
\end{equation}
where, $G \equiv \exp\{- \frac{1}{2} (x_1^2 + x_2^2 + x_3^2)\}$, $X \equiv
x_{12} x_{23} x_{31}$, $Y \equiv y_{12} y_{23} y_{31}$, $g =
\alpha (\alpha - 1)$ and $f = \lambda (\lambda - 1)$. Now, one can perform the
following ST:
\begin{equation}
\tilde{H} \equiv \psi_0^{-1} H \psi_0 = 2 \sum_{i=1}^3 x_i \partial_i -
\hat{A} + E_0\qquad,
\end{equation}
where, $E_0 \equiv 3 (2 \alpha + 2 \lambda +1)$ and $\hat{A} \equiv
\sum_{i=1}^3 {\partial_i}^2 + 2 \alpha \sum_{i<j}
\frac{1}{x_{ij}} \hat{d}_{ij} + 2 \lambda \sum_{i<j} \frac{1}{y_{ij}}
\hat{D}_{ij}$; here $\hat{d}_{ij} \equiv {\partial}_i - {\partial}_j$ and
$\hat{D}_{ij} \equiv {\partial}_i + {\partial}_j - 2 {\partial}_k$.

Since the Euler operator $\sum_i x_i \partial_i$ measures the degree of any
homogeneous function of $x_i$ and $\partial_i$, it is easy to verify the
commutation relation 
\begin{equation}
[2 \sum_i x_i \partial_i\,\,,\,\,\exp\{- \frac{1}{4} \hat{A}\}] =
\hat{A}\,\,\exp\{- \frac{1}{4} \hat{A}\}\qquad.
\end{equation}
The above result can be used to make another ST on $\tilde{H}$, which yields
\begin{equation}
\bar{H} \equiv \exp\{\frac{1}{4} \hat{A}\}\,\,\tilde{H}\,\,\exp\{- \frac{1}{4}
\hat{A}\} = 2 \sum_i x_i \partial_i + E_0\qquad.
\end{equation}
One more ST on $\bar{H}$ by $\hat{T} \equiv \exp\{\frac{1}{4} \sum_i
\partial_i^2\}\,\,G^{-1}$ establishes the connection of the CMW Hamiltonian
with those of free harmonic oscillators, {\it i.e.},
\begin{equation}
\hat{T}^{-1}\,\,\bar{H}\,\,\hat{T} = \sum_{i=1}^3 (- \partial_i^2 + x_i^2) +
(E_0 - 3)\qquad.
\end{equation}
It is worth mentioning that for the normalizability of the wavefunctions, one
needs to check that the action of $\exp\{- \hat A/4\}$ on an appropriate
linear combination of the eigenstates of $\sum_{i=1}^3 x_i
\partial_i$ yields a polynomial solution. We observe that the above example
is an explicit realization of the more general result established earlier in
[9]: all $D$ dimensional $N$ particle Hamiltonians which can be brought
through a suitable transformation to the generalized form, $\tilde H =
\sum_{l=1}^D \sum_{i=1}^N x_i^{(l)} \frac{\partial}{\partial x_i^{(l)}} + E_0
+ \hat A$ can also be mapped to $\sum_{l=1}^D \sum_{i=1}^N x_i^{(l)}
\frac{\partial}{\partial x_i^{(l)}} + E_0$ by $\exp\{- d^{-1} \hat A\}$; where,
the operator $\hat A$ is any homogeneous function of
$\frac{\partial}{\partial x_i^{(l)}}$ and $x_i^{(l)}$ with degree $d$ and
$E_0$ is a constant. It should be emphasized that, this procedure reproduces
the linear part of the spectrum of the original Hamiltonian.

For the purpose of constructing the eigenfunctions, one can make use of (5).
It is interesting to note that, although the monomials $\prod_{i=1}^3
x_i^{m_i}$ are the eigenfunctions of $\bar{H}$ with eigenvalues
$E_{m_1,m_2,m_3} \equiv 2 (m_1 + m_2 + m_3 ) + E_0$, they are not acceptable
as the eigenfunctions of the CMW Hamiltonian because, the action of $\exp\{-
\frac{1}{4} \hat{A}\}$ on them do not yield polynomial solutions. However, the
powers of the symmetric combinations, $R \equiv \frac{1}{3} \sum_{i=1}^3
x_i$, $r^2 \equiv \frac{1}{3} \sum_{i<j} x_{ij}^2 = \frac{1}{9} \sum_{i<j}
y_{ij}^2$ and $Y \equiv y_{12} y_{23} y_{31}$ in the form
$R^{n_1}\,\,(r^2)^{n_2}\,\,Y^{2 n_3}$, are not only the eigenfunctions of
$\bar{H}$ with eigenvalues $E_{n_1,n_2,n_3} \equiv 2 (n_1 + 2 n_2 + 6 n_3) +
E_0$ but also yield polynomial solutions upon the action of $\exp\{-
\frac{1}{4} \hat{A}\}$; here, $n_1, n_2, n_3 = 0, 1, 2, \cdots$. Therefore,
the normalizable wavefunctions of the CMW  Hamiltonian $H$, in the Cartesian
basis, are given by 
\begin{equation}
\psi_{n_1,n_2,n_3} = \psi_0\,\,\exp\{- \frac{1}{4} \hat{A}\}
\left(R^{n_1}\,\,(r^2)^{n_2}\,\,Y^{2 n_3}\right)\qquad.
\end{equation}

In the following, we explicitly construct some unnormalized eigenfunctions of
(1) using (7). 

\noindent {\it Case I}: $n_2 = n_3 = 0$

This case corresponds to center-of-mass degree of freedom:
\begin{equation}
\psi_{n_1,0,0} = \psi_0\,\,\exp\{- \frac{1}{4} \hat{A}\}
R^{n_1} = \exp\{- \frac{1}{4} \sum_{N=1}^3 \partial_i^2\} \,\,R^{n_1}\qquad.
\end{equation}
This can be cast in the form [11],
\begin{equation}
\psi_{n_1,0,0} = 6^{- n_1} n_1! \psi_0\,\,\sum_{\sum_{i=1}^3 m_i = n_1}
\prod_{i=1}^3 \frac{H_{m_i}(x_i)}{m_i!} \qquad.
\end{equation}
Here, $H_{m_i}(x_i)$ are the Hermite polynomials.

\noindent {\it Case II}: $n_1 = n_3 = 0$

Another orthogonal set characterized by the quantum number $n_2$ can be
written in the form,
\begin{equation}
\psi_{0,n_2,0} = \psi_0\,\,\exp\{- \frac{1}{4} \hat{A}\}
\,\,(r^2)^{n_2}\qquad.
\end{equation}
It can easily be checked that $\hat{A} {(r^2)}^n = 4 n (n + 3 \alpha + 3 \lambda)
{(r^2)}^{n-1}$; this gives $\psi_{0,{n_2},0}$ as 
\begin{eqnarray}
\psi_{0,n_2,0} &=& \psi_0\,\,(-1)^{n_2} n_2! \sum_{m=0}^{n_2} \frac{(-
1)^m}{m! (n_2 - m)!} \frac{(3\alpha + 3\lambda + n_2)!}{(3\alpha + 3\lambda + 
m)!} {(r^2)}^{m} \qquad,\nonumber\\ 
&=& \psi_0\,\,(-1)^{n_2}\,\,n_2! \,\,L_{n_2}^{3\alpha + 3\lambda}(r^2) \qquad.
\end{eqnarray}
Here, $L_{n_2}^{3\alpha + 3\lambda}(r^2)$ is the Lagurre polynomial. 

\noindent {\it Case III}: $n_1 = n_2 = 0$ 

Now, (7) becomes
\begin{equation}
\psi_{0,0,n_3} = \psi_0\,\,\exp\{- \frac{1}{4} \hat{A}\}
Y^{2 n_3}\qquad.
\end{equation}
We note that, on the odd powers of $Y$, the
action of $\exp\{- \frac{1}{4} \hat{A}\}$ does not yield polynomial solutions
and hence the resulting states are not normalizable. 
As an example,
\begin{equation}
\hat{A} Y = 2 \lambda \frac{1}{Y} (y_{12}^4 + y_{23}^4 + y_{31}^4) \qquad.
\end{equation}
It is clear that the action of $\exp\{- \frac{1}{4} \hat{A}\}$ on $Y$ will
contain negative powers of $Y$ and can not be terminated as a polynomial.
The wavefunctions can be computed by making use of 
\begin{equation}
\hat{A} Y^{2 n_3} = 4 n_3 (3 [2 n_3 - 1] + \lambda)
Y^{2 (n_3 - 1)} (y_{12}^4 + y_{23}^4 + y_{31}^4)\qquad.
\end{equation}
As an example, the $n_3 = 1$ state is given by
\begin{equation}
\psi_{0,0,1} = \psi_0\,\,\left(Y^2 - (3 + \lambda) \sum_{i<j}^3 y_{ij}^4 -
\frac{9}{2} (2 + 3 \alpha + 3 \lambda) \sum_{i<j}^3 y_{ij}^2 - \frac{3}{4} (2 + 3
\alpha + 3 \lambda)^2 \right) \qquad.
\end{equation}
It should be noted that, in the Jacobi coordinates, the above set of
wavefunctions involve angle variable and hence is orthogonal to the former
two sets. 

The underlying algebraic structure of CMW model can also be found easily from
(5), by defining the creation and annihilation operators as $a_i^+ = \hat S
x_i {\hat S}^{-1}$ and $a_i^- = \hat S \partial_i {\hat S}^{-1}$:
$[a_i^-\,\,,\,\,a_j^+] = \delta_{ij}$ and the CMW Hamiltonian becomes
\begin{equation}
H = 2 \sum_i H_i = \sum_i \{a_i^-\,\,,\,\,a_i^+\} + (E_0 - 3)
\qquad,
\end{equation}
where, $\hat S \equiv \psi_0\,\,\exp\{- \frac{1}{4} \hat A\}$, $H_i \equiv
\frac{1}{2} \{a_i^-\,\,,\,\,a_i^+\} + \frac{1}{6} (E_0 - 3)$ and
$[H_i\,\,,\,\,a_i^- (a_i^+)] = - a_i^- (a_i^+)$. Here, we would like to
remark that, the states created by the action of individual $a_i^+$ on the
ground-state $|0>$ which is obtained from $a_i^- |0> = 0$, are not
normalizable unlike their free counterparts; however their symmetric
combinations (7) are found to be normalizable. This shows that there are no
single particle excitations and any excited state will contain all the three
particles in some symmetric state. In other words, the present analysis gives
an algebraic statement about a truly correlated system.

Now, the integrability of the CMW model can be seen easily. It is obvious
that $[H\,\,,\,\,H_k] = [H_i\,\,,\,\,H_j] = 0; i,j,k = 1,2,3$. Therefore, the
set $\{H_1, H_2, H_3\}$ provides the three conserved quantities. From this
set, one can construct, three linearly independent symmetric conserved
quantities.  This proof of integrability is entirely different from the one
given in [8].  It is of considerable interest to note that, all the above
analyses done for the CMW model will go through even in the absence of pure
two-body terms ($\alpha = 0$ or $1$), {\i.e.}, a model with pure three-body
inverse-square interaction also shares the same algebraic structure as that
of the original CMW model. It is worth mentioning that, in the limit $g
\rightarrow 0$ or $f \rightarrow 0$, CMW model reproduces only a part of the
spectrum of the Hamiltonians with pure two-body or three-body inverse-square
potentials respectively. This happens due to the presence of the singular
interactions [1].

The above technique can be extended to a one parameter family of potentials
connected to the CMW model [12]. The Hamiltonian is given by
\begin{equation}
H = - \sum_{i=1}^3 \partial_i^2 + \sum_{i=1}^3  x_i^2
+ g \sum_{i\ne j}^3 \frac{1}{{(x_{ij} \cos\delta +
+ \frac{1}{\sqrt{3}} y_{ij} \sin\delta)}^2}\qquad,
\end{equation}
where, $0 \le \delta \le \pi/6$.
Performing a ST on the above Hamiltonian by the ground-state wavefunction
$\psi_0$, one gets 
\begin{equation}
H^\prime \equiv {\psi_0}^{-1} H \psi_0 = 2 \sum_i^3 x_i \partial_i 
- \hat B + E_0\qquad,
\end{equation}
where, $E_0 = 3 + 6 \alpha$, $G$ and $\alpha$ as before, $\psi_0 = G
X^\alpha$, $X \equiv \prod_{i<j}^3 X_{ij}$, $X_{ij} \equiv {[(x_i - x_j)
\cos\delta + \frac{(x_i + x_j - 2 x_k)}{\sqrt{3}} \sin\delta]}$ and $\hat B
\equiv \sum_{i=1}^3 \partial_i^2 + 2 \alpha \sum_{i<j}^3
\frac{1}{X_{ij}} {\hat D}_{ij}$; here, ${\hat D}_{ij} \equiv (\cos\delta +
\frac{1}{\sqrt{3}}\sin\delta) \partial_i + (- \cos\delta +
\frac{1}{\sqrt{3}}\sin\delta) \partial_j - (\frac{2}{\sqrt{3}} \sin\delta)
\partial_k$ and $i\ne j \ne k \ne i$. It should be noted that $\psi_0$
interpolates smoothly between the pure two-body and three-body cases for
$\delta = 0$ or $\delta = \pi/6$ respectively.

Analogous to (3), (18) can also be mapped to free oscillators, {\it i.e},
\begin{equation}
\tilde{H} \equiv \exp\{\frac{1}{4} \hat B\}\,\,H^\prime\,\,\exp\{-
\frac{1}{4} \hat B\} = 2 \sum_i^3 x_i \partial_i + E_0 \qquad, 
\end{equation}
and
\begin{equation}
{\hat{T}}^{-1}\,\,\tilde{H}\,\, \hat{T} = - \sum_{i=1}^3
\partial_i^2 + \sum_{i=1}^3 x_i^2 + (E_0 - 3) \qquad,
\end{equation}
where, $\hat{T}$ is as given earlier. By inverse ST, one can recast the above
Hamiltonian in the form of decoupled oscillators as given in (16).
Construction of the eigenfunctions and proving the quantum integrability of
this model can be carried out in parallel to the CMW model. Similar analyses
can also be extended to more generalized potentials of the above form [12].

We list below other non-trivial, interacting, three-body potentials with
linear spectra [7], 
$$V(x_1, x_2, x_3) = g \sum_{i<j} x_{ij}^{-2} - \frac{f_3}{\sqrt{6} r}
\sum_{i<j} \frac{y_{ij}}{x_{ij}^2}\qquad,$$
and
$$V(x_1, x_2, x_3) = 3 f \sum_{i<j} y_{ij}^{-2} + \frac{f_3}{3 \sqrt{2} r}
\sum_{i<j} \frac{x_{ij}}{y_{ij}^2}\qquad,$$ 
where, $r$, $x_{ij}$ and $y_{ij}$ are as given earlier. One should note that
these potentials contain the variable $r$ explicitly and hence the
corresponding quantum mechanical problems can be better tackled in the polar
coordinates. We have checked that these models can also be made equivalent to
free oscillators.

Finally, we present an example involving four particles, to demonstrate the
usefulness of our method for finding new solvable models. Since, these models
can be mapped to a set of free oscillators, their eigenspectra are guaranteed
to be linear. The Hamiltonian reads,
\begin{eqnarray}
H &=& - \sum_{i=1}^4 {\partial}_i^2
+ \sum_{i=1}^4 x_i^2 + \alpha (\alpha - 1) \sum_{{i,j=1}\atop {i\ne j}}^4
\frac {1}{x_{ij}^2} + 2 \lambda (\lambda - 1) \sum_{{i,j,k=1}\atop {{{i\ne
j}\ne k}\ne i}}^4 \frac {1}{y_{ijk}^2} \nonumber\\
&+& \frac{2}{3} \lambda (\lambda + 4
\alpha) \frac{1}{Y} \sum_{{i,j,k=1}\atop {{{i\ne j}\ne k}\ne i}}^4 y_{ijk}^2
\,\,, 
\end{eqnarray}
here, $x_{ij}$ is as before and $Y \equiv y_{123} y_{234} y_{341} y_{412}$
with, $y_{ijk} \equiv x_i + x_j + x_k - 3 x_l$. Here, we would like to
remark that, the four-particle system, with pure inverse-square two-body
interactions, is integrable, whereas the one with pure three-body interaction
is not.

The ground-state wavefunction is given by
\begin{equation}
\psi_0 = \exp\{- \frac{1}{2} \sum_{i=1}^4 x_i^2\}\,\,\prod_{i<j}|x_i -
x_j|^\alpha\,\,|Y|^\lambda \qquad.
\end{equation}

The equivalence of the above Hamiltonian to a set of free oscillators follows
from 
\begin{equation}
(\psi_0\,\,\exp\{- \frac{1}{4}
\hat{C}\}\,\,\hat{T})^{-1}\,\,H\,\,(\psi_0\,\,\exp\{- \frac{1}{4}
\hat{C}\}\,\,\hat{T}) = - \sum_{i=1}^4 \partial_i^2 + \sum_{i=1}^4 x_i^2 + (E_0
- 4) \qquad,
\end{equation}
where, $\hat{C} \equiv \sum_{i=1}^4 \partial_i^2 + 2 \alpha \sum_{i < j}^4
\frac{1}{x_{ij}} {\hat{d}}_{ij} + 2 \lambda \sum_{i<{j<k}} \frac{1}{y_{ijk}}
{\hat{F}}_{ijk}$; ${\hat{F}}_{ijk} \equiv \partial_i + \partial_j +
\partial_k - 3 \partial_l$ and $E_0 = 4 + 8 \alpha + 8 \lambda$. $\hat{T}$,
$x_{ij}$ and ${\hat{d}}_{ij}$ are similar to those given earlier.  In
parallel to the CMW model, one can construct eigenfunctions, show the
harmonic oscillator algebra and prove the quantum integrability for this
model. As is clear, this method can be extended to $N$ particle systems.
Generalization of this model analogous to the one given in (17) can also be
dealt in the same manner.

In conclusion, we have shown that a number of models having complicated
few-body interactions but with linear energy spectra like harmonic
oscillators, can indeed be made equivalent to free oscillators by similarity
transformations. There are no single particle excitations in all these
models; the wavefunctions contain all the particles in some symmetric
combination.  Although the underlying algebraic structure of all these models
is that of free harmonic oscillators, unlike the oscillator case, the
individual states generated by the creation operators are not normalizable.
Only their symmetric combinations are normalizable. Our analysis gave an
algebraic statement about a truly correlated system. We conjecture that, all
the correlated physical systems in nature will have these features of the CMW
model; however, the underlying algebraic structure may be different from that
of decoupled harmonic oscillators.

Our method allows one to prove the quantum integrability in a straightforward
manner and to construct new interacting solvable models.  It is amusing to
note that the ground-states of these models are similar to the ones that
describe edge excitations in quantum Hall effect [13]. Since the planar
wavefunctions have exact correspondence with their one dimensional
counterparts [14], our technique can be of potential use for these physical
systems. Extension to other many-body Hamiltonians will also through light on
the structure of these complicated interacting systems [15,16]. Finally, we
conjecture that, any $N$-body problem having a linear eigenvalue spectrum can
be reduced to a set of decoupled harmonic oscillators by some suitable
transformation.

P.K.P would like to acknowledge useful discussions with Profs. V. Srinivasan
and S. Chaturvedi and Dr. M. Sivakumar.  N.G thanks U.G.C (India) for
financial support through the S.R.F scheme.


\begin{references}
\item F. Calogero, J. Math. Phys. {\bf 10}, 2191 (1969).
\item F. Calogero, J. Math. Phys. {\bf 10}, 2197 (1971); {\bf
      12}, 419 (1971).
\item B. Sutherland, J. Math. Phys. {\bf 12}, 246 (1971); {\bf 12}, 251
      (1971). 
\item For various connections, see the chart in B.D. Simons, P.A. Lee and
      B.L. Altshuler, Phys. Rev. Lett. {\bf 72}, 64 (1994).
\item J. Wolfes, J. Math. Phys. {\bf 15}, 1420 (1974).
\item F. Calogero and C. Marchioro, J. Math. Phys. {\bf 15}, 1425 (1974).
\item A. Khare and R.K. Bhaduri, J. Phys. {\bf A 27}, 2213
      (1994). 
\item C. Quesne, Mod. Phys. Lett. {\bf A 10}, 1323 (1995).
\item N. Gurappa and P.K. Panigrahi, cond-mat/9710035.
\item N. Gurappa and P.K. Panigrahi, quant-ph/9710019.
\item K. Vacek, A. Okiji and N. Kawakami, J. Phys. {\bf A 29}, L201 (1994); 
      N. Gurappa and P.K. Panigrahi, Mod. Phys. Lett. {\bf A 11}, 891 (1996).
\item A. Khare and U.P. Sukhatme, pre-print/IOP-BBSR/97-38, quant-ph/9802050,\\
      to appear in Phys. Lett. {\bf A}.
\item M. Stone, Quantum Hall Effect (World Scientific, 1992) and references
      therein.
\item H. Azuma and S. Iso, Phys. Lett. {\bf B 331}, 107 (1994); 
      P.K. Panigrahi and M. Sivakumar, Phys. Rev. {\bf B 52}, 13742 (15)
      (1995). 
\item N. Gurappa, C.N. Kumar and P.K. Panigrahi, Mod. Phys. Lett. {\bf A 11},
      1737 (1996). 
\item A. Khare, cond-mat/9712133.
\end{references}
\end{document}